# Structure and electronic structure of Metal-Organic Frameworks within the Density-Functional based Tight-Binding method

**Binit Lukose,**[1] **Barbara Supronowicz,**[1] **Petko St. Petkov,**[2] **Johannes Frenzel,**[3] **Agnieszka B. Kuc,**[1] **Gotthard Seifert,**[4] **Georgi N. Vayssilov,**[2] **Thomas Heine**[*,1]

[1] School of Engineering and Science, Jacobs University Bremen, Campus Ring 1, 28759, Bremen Germany
[2] Department of Chemistry, University of Sofia, Sofia, Bulgaria
[3] Lehrstuhl für Theoretische Chemie, Ruhr-Universität Bochum 44780 Bochum, Germany
[4] Physikalische Chemie, Technische Universität Dresden, Bergstrasse 66b, 01062 Dresden, Germany

**Dedicated to the 60th birthday of Prof. Thomas Frauenheim**

Density-functional based tight-binding is a powerful method to describe large molecules and materials. Metal-Organic Frameworks (MOFs), materials with interesting catalytic properties and with very large surface areas have been developed and have become commercially available. Unit cells of MOFs typically include hundreds of atoms, which make the application of standard Density-Functional methods computationally very expensive, sometimes even unfeasible. The aim of this paper is to prepare and to validate the Self-Consistent Charge Density-Functional based Tight Binding (SCC-DFTB) method for MOFs containing Cu, Zn and Al metal centers. The method has been validated against full hybrid density-functional calculations for model clusters, against gradient corrected density-functional calculations for supercells, and against experiment. Moreover, the modular concept of MOF chemistry has been discussed on the basis of their electronic properties. We concentrate on MOFs comprising three common connector units: copper paddlewheels (HKUST-1), zinc oxide $Zn_4O$ tetrahedron (MOF-5, MOF-177, DUT-6 (MOF-205)) and aluminium oxide $AlO_4(OH)_2$ octahedron (MIL-53). We show that SCC-DFTB predicts structural parameters with a very good accuracy (with less than 5% deviation, even for adsorbed CO and $H_2O$ on HKUST-1), while adsorption energies differ by 12 kJ mol$^{-1}$ or less for CO and water compared to DFT benchmark calculations.

## 1 Introduction

In reticular, or modular chemistry, molecular building blocks are stitched together to form regular frameworks [1]. With this concept it became possible to construct framework compounds with interesting structural and chemical composition, most notably metal-organic frameworks (MOFs) [1-10] and covalent organic frameworks (COFs) [11-17]. The interest in MOFs and COFs is not limited to chemistry, these crystalline materials are also interesting for applications in information storage [18], sieving of quantum liquids 19], hydrogen storage [20], and for fuel cell membranes [21-23].

COF and MOF frameworks are composed by combining two types of building blocks, so-called connectors, typically coordinating in 4-8 sites, and linkers, which have typically 2, sometimes also 3 or 4 connecting sites. Fig. 1 illustrates an simplified representation of the topology of connectors and linkers, by using secondary building units (SBUs) (see Fig. 1).

Linkers are organic molecules, with carboxylic acid groups at their connection sites, which form bonds to the connectors (typically in a solvothermal condensation reaction). They can carry functional groups, which can make them interesting for applications in catalysis [24]. Connectors are either metal organic units, as for example the well-known $Zn_4O(CO_2)_6$ unit of MOF-5 [8], the $Cu_2(HCOO)_4$ paddle wheel unit of HKUST-1 [10], or covalent units incorporating boron oxide linking units for COFs [11]. As the building blocks of MOFs and COFs comprise typically some 10 atoms, unit cells quickly get very large. In particular if adsorption or dynamic processes in MOFs and COFs are of interest, (super)cells of some 1000 atoms need to be processed. While standard organic force fields show a reasonable performance for COFs [25], the creation of reliable force fields is not straight-forward for MOFs as transferable parameterization of the transition metal sites are an issue, even though progress has been achieved for selected materials [26,27]. The difficulty to describe transition metals, in particular if they are catalytically active as in HKUST-1, limits the applicability of molecular mechanics (MM) even for QM/MM hybrid methods[28].

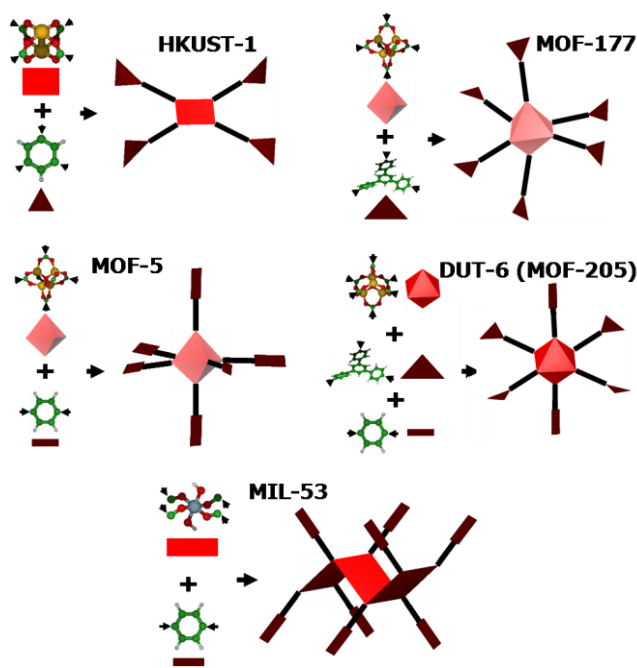

**Figure 1** The connector and the linker of the frameworks CuBTC (HKUST-1), MOF-177, MOF-5, DUT-6 (MOF-205) and MIL-53 and their respective secondary-building units (SBUs). The arrows indicate the connection points of the fragments. Red-oxygen, green-carbon, white-hydrogen, yellow-copper/zinc, blue-aluminum.

On the other hand, the density-functional based tight-binding (DFTB) method with its self-consistent charge (SCC) extension to improve performance for polar systems is a computationally feasible alternative. This non-orthogonal Tight-Binding approximation to density-functional theory (DFT), which has been developed by Seifert, Frauenheim, and Elstner and their groups [29-32] (for a recent review see Ref. 30), and which has been successfully applied to a large scale systems such as biological molecules [33-36], supramolecular systems [37,38], surfaces [39,40], liquids and alloys [41,42] and solids [43]. Being a quantum-mechanical method and hence allowing the description of breaking and formation of chemical bonds, the method showed outstanding performance in the description of processes such as the mechanical manipulation of nanomaterials [44-46]. It is remarkable that the method performs well for systems containing heavier elements such as transition metals, as this domain cannot be covered so far with acceptable accuracy by traditional semi-empirical methods [47,48]. DFTB covers today a large part of the elements of the periodic table, and parameters and a computer code are available from the DFTB.org website.

Recently, we have validated DFTB for its application for COFs [16,17], and have shown that structural properties and formation energies of COFs are well-described within DFTB. Kuc et al. have validated DFTB for substituted MOF-5 frameworks, where connectors are always the $Zn_4(CO_2)_6$ unit, which has been combined with a large variety of organic linkers [49].

In this work, we have revised the DFTB parameters developed for materials science applications and validated them for HKUST-1 and, being far more challenging, for the interaction of its catalytically active Cu sites with carbon monoxide and water. The $Cu_2(HCOO)_4$ paddle wheel units are electronically very intriguing: On a first note, the electronic ground state of the isolated paddle wheel is an antiferromagnetic singlet state, which cannot be described by one Slater determinant and which is consequently not accessible for Kohn-Sham DFT. However, the energetically very close triplet state correctly describes structure and electronic density of the system, and also adsorption properties agree well with experiment [50-52]. We therefore use DFT in the B3LYP hybrid functional representation as benchmark for DFTB validations for HKUST-1, added by DFT-GGA calculations for periodic unit cells. We will show that the general transferability of the DFTB method will allow investigating structural, electronic and in particular dynamic properties.

**2 Computational Details**

All calculations of the finite model and periodic crystal structures of MOFs were carried out using the dispersion-corrected self-consistent density functional based tight-binding (DC-SCC-DFTB, in short DFTB) method [30,53-55] as implemented in deMonNano code [56]. Two sets of parameters have been used: the standard SCC-DFTB parameter set developed by Elstner et al. [54] for Zn-containing MOFs, for Cu- and Al-containing materials we have extended our materials science parameter set, which has been developed originally for zeolite materials to include Cu. For this element we have used the standard procedure of parameter generation: we have used the minimal atomic valence basis for all atoms including polarization functions when needed. Electrons below the valence states were treated within the frozen-core approximation. The matrix elements were calculated using the local density approximation (LDA), while the short-range repulsive pair-potential was fitted to the results from DFT general gradient approximation (GGA) calculations. For more details on DFTB parameter generation see Ref.[30].

The reference DFT calculations of the equilibrium geometry of the investigated isolated MOF models were performed with density functional theory employing the Becke three-parameter hybrid method combined with a LYP correlation functional (B3LYP)[57,58]. The basis set associated with the Hay-Wadt [59] relativistic effective core potentials proposed by Roy et al. [60] was supplemented with polarization $f$ functions [61] and was employed for description of the electronic structure of Cu atoms. The Pople 6-311G(d) basis sets were applied for the H, C and O atoms. The calculations were performed with the Gaussian09 program suite [62]. For optimization of the periodic structure of CuBTC at DFT level the electronic structure code Quickstep[63], which is part of the CP2K package [64], was used. Quickstep is an implementation of the Gaussian plane wave method [65], which is based on the Kohn-Sham formulation of density functional theory (DFT). We have used the generalized gradient functional PBE [66] and the Goedecker-Teter-Hutter pseudopotentials [67,68] in conjunction with double-ζ basis sets with polarization functions DZVP-MOLOPT-SR-GTH basis sets obtained as described in Ref [69] but using two less diffuse primitives. The plane wave basis with cutoff energy of 400 Ry was used throughout the simulations.

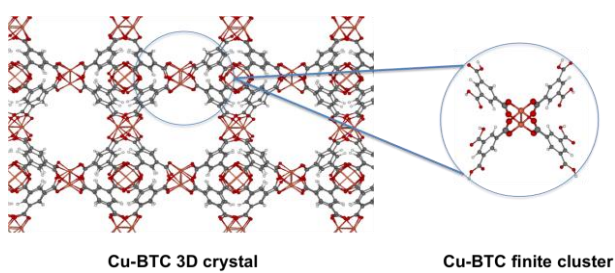

**Figure 2** Crystal structure and cluster model of Cu-BTC MOF as used in the computer simulations.

The initial crystal structures were taken from experiment (powder X-ray diffraction (PXRD) data). The cell parameters and the atomic positions were fully optimized using conjugate gradient method at the DFTB level. For the reference DFT calculations, only the atomic positions of experimental crystal structures were minimized. The cluster models were cut from the optimized structures and saturated with hydrogen atoms (see Fig. 2 for exemplary systems of Cu-BTC).

### 3 Results and Discussion

**3.1 Cu-BTC** We have optimized the crystal structure of Cu-BTC (HKUST-1) [10] using cluster models and the periodic crystal structure. The structural properties were compared to DFT results (see Table 1). The geometries were obtained in the gas phase and in the presence of water molecules (see Table 2), as the synthesized crystals often have $H_2O$ molecules attached to the open metal sites of Cu. We have also studied changes of the cluster model geometry in the presence of carbon monoxide (CO) and the results are also shown in Table 2. We have obtained good agreement with experimental data as well as with DFT results.

**Table 1** Selected bond lengths [Å] and bond angles [°] of Cu-BTC optimized at DFTB and DFT level. Cluster models and periodic crystal structures are compared in the gas phase and in the presence of water molecules.

| Bond Type | Cluster Model | Crystal Structure | Exp. |
|---|---|---|---|
| Cu-Cu | 2.50 (2.57) | 2.50 (2.50) | 2.63 |
| Cu-O | 2.05 (1.98) | 2.02 (1.98) | 1.95 |
| O-C | 1.34 (1.33) | 1.33-1.38 (1.28) | 1.25 |
| OCuO | 83.6-97.1 (89.8) | 89.2-90.7 (87.3-93.7) | 89.1, 89.6 |
| Cell paramet. | | a=b=c=27.283 (26.343) α=β=γ=90 (90) | a=b=c=26.343 α=β=γ=90 |

**Table 2** Selected bond lengths [Å] and bond angles [°] of Cu-BTC optimized at DFTB and DFT level in the presence of water molecules. Cluster models are calculated using water and carbon monoxide molecules. The adsorption energies $E_{ads}$ of the guest molecules are given in [kJ mol$^{-1}$].

| Bond Type | Cluster Model + $H_2O$ | Crystal Structure + $H_2O$ | Cluster Model + CO |
|---|---|---|---|
| Cu-Cu | 2.67 (2.66) | 2.62 (2.60) | 2.50 (2.60) |
| Cu-O | 2.05 (1.97-2.06) | 2.10 (1.96-2.00) | 2.06 (1.99) |
| O-C | 1.34 (1.27) | 1.33 (1.28) | 1.34 (1.27) |
| OCuO | 84.3-95.5 (88.9-90.5) | 87.1-92.1 (84.2-93.0) | 84.2-96.7 (89.6) |
| Cu-O(H2O) Cu-C(CO) | 2.37 (2.19) | 2.44 (2.33-2.55) | 3.07 |
| $E_{ads}$ | -40.45 (-52.00) | | -16.48 (-28.00) |

We have also simulated the XRD patterns of Cu-BTC, which show very good structural agreement between the experimental and calculated structures (see Fig. 3).

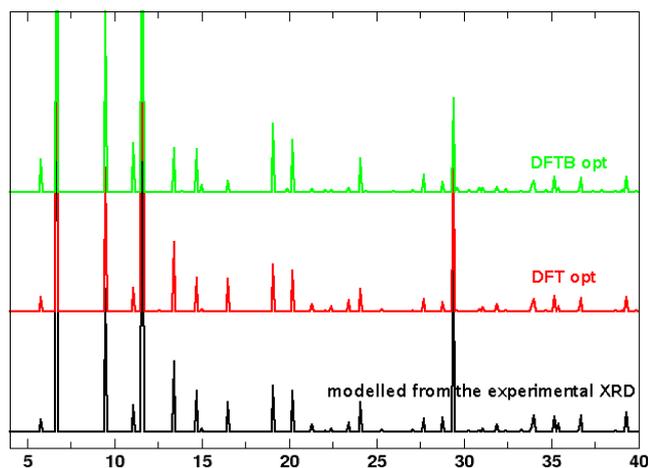

**Figure 3** Simulated XRD patterns of Cu-BTC MOF with the experimental unit cell parameters and optimized at the DFTB and DFT level of theory.

In detail: The bond lengths and bond angles do not change significantly when going from the cluster model to the crystal structure, confirming the reticular chemistry approach. The only exception is the O-Cu-O bond angle that differs by 4.-7° between the two systems at both levels of theory.

The bond length between Cu atoms is slightly underestimated comparing with experimental (by maximum 5%) and DFT (maximum by 3%) results, while all other bond lengths are somewhat larger at DFTB.

All bond lengths stay unchanged or become longer in the presence of water molecules. The most striking example is the Cu-Cu bond, where the change reaches 0.12-0.17 Å compared to the gas phase. Also, bond angles increase by up to 2˚, when the $H_2O$ is present. The Cu-O bond length with the oxygen atom from the carboxylate groups is shorter than that with the oxygen atoms from the water molecules, indicating that the latter is only weakly bound to the copper ions. The O-C distances (~1.33 Å) correspond to values between that for the typical single (1.42 Å) and double (1.22 Å) oxygen-carbon bond. The calculated $C_{ring}$-$C_{carboxyl}$ bond lengths of 1.46 Å indicate that these are typical single $sp^2$-$sp^2$ C-C bonds, while experimental findings give a slightly longer value (1.5 Å) for this MOF.

The unit cell parameters with and without water molecules obtained at the DFTB level overestimate the experimental data by less than 4%, which gives a fairly good agreement if we take into account high porosity of the material. These lattice parameters increase only slightly from 27.283 to 27.323 Å in the presence of water.

We have calculated the binding energies of $H_2O$ molecules attached to the Cu-metal sites within the cluster model. At the DFTB level, $E_{bind}$ of 40 kJ mol$^{-1}$ was obtained in a good agreement with the DFT results (52 kJ mol$^{-1}$). We have also calculated the binding energy of CO, which was twice smaller than that of $H_2O$ (16.5 and 28 kJ mol$^{-1}$ for DFTB and DFT, respectively), suggesting much stronger binding of O atom (from $H_2O$) than C atom (from CO). While the bond lengths in the MOF structure do not change in the presence of either $H_2O$ or CO, the differences in the binding energy come from much longer bond distances (by around 0.7 Å) for Cu-C than for Cu-O in the presence of carbon monoxide and water molecules, respectively.

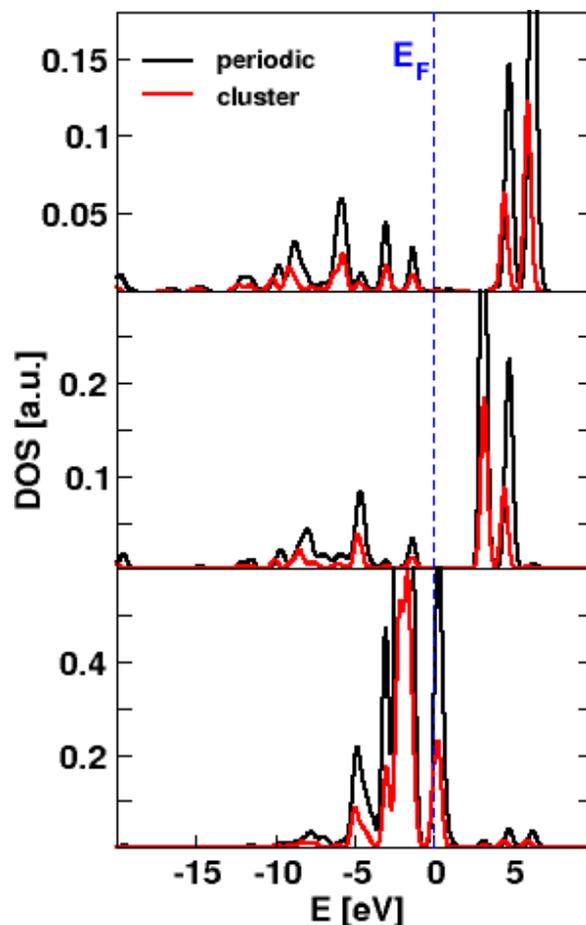

**Figure 4** PDOS of Cu (top) *s*-orbitals, (middle) *p*-orbitals and (bottom) *d*-orbitals of Cu-BTC in the form of crystal and cluster structures.

Furthermore, we have studied the electronic properties of periodic and cluster model Cu-BTC by means of partial density of states (PDOS). We have compared especially the Cu-PDOS for *s*-, *p*- and *d*-orbitals (see Fig. 4). The results show that the electronic structure stays unchanged when going from the cluster models to the fully periodic crystal of Cu-BTC. Since the copper atoms are of Cu(II), the *d*-orbitals are just partially occupied, what results in the metallic states at the Fermi level. This is a very interesting result, as other Zn-O-based MOFs are either semiconductors or insulators.[49]

### 3.2 MOF-5, -177, DUT-6 (MOF-205) and MIL-53

We have also studied the structural properties of MOF

structures with large surface area using the SCC-DFTB method. Very good agreement with the experimental data shows that this method is applicable for such structurally very diverse structures, as well as for coordination polymers based on the MOF-5 framework, which has been reported earlier [49]. Tables 3 and 4 show selected bond lengths and bond angles of MOF-5 [70,] MOF-177 [71], DUT-6 (MOF-205) [72,73] and MIL-53 [74], respectively.

**Table 3** Selected bond lengths [Å] and bond angles [°] of MOF-5 (424 atoms in unit cell), MOF-177 (808 atoms in unit cell) and DUT-6 (MOF-205) (546 atoms in unit cell) optimized at DFTB level. The available experimental data is given in parenthesis.[70-73] O' denotes the central O atom in the $Zn-O$-octahedron.

| Bond Type | MOF-5 | MOF-177 | DUT-6 (MOF-205) |
|---|---|---|---|
| Zn-Zn | 3.30 (3.17) | 3.22-3.36 (3.06-3.30) | 3.25-3.31 (3.18) |
| Zn-O' | 2.02 (1.94) | 2.02 (1.93) | 2.02 (1.94) |
| Zn-O | 2.04 (1.92) | 2.02-2.06 (1.90-1.99) | 2.02, 2.05 (1.93) |
| O-C | 1.28 (1.30) | 1.28 (1.31) | 1.28 (1.25) |
| ZnO'Zn | 109.5 (109.5) | 105.6-112.4 (105.5, 109.2) | 107-111.8 (108.4, 110.0) |
| OZnO | 108.3, 110.8 (106.1) | 104.8, 114.5 (98.1-128.1) | 104.6-111.2 (106.2, 108.5) |
| Cell paramet. | a=b=c=26.472 (25.832) $\alpha=\beta=\gamma=90$ (90) | a=b=37.872 (37.072) c=30.68 (30.033) $\alpha=\beta=90$ (90) $\gamma=120$ (120) | a=b=c=31.013 (30.353) $\alpha=\beta=\gamma=90$ (90) |

MOFs-5, -177 and DUT-6 (MOF-205) are built of the same connector, which is $Zn_4O(CO_2)_6$ octahedron. The difference is in the organic linker: 1,4-benzenedicarboxylate (BTC), 1,3,5- benzenetibenzoates (BTB) and BTB together with 2,6-naphtalenedicarboxylate (NDC) for MOF-5, -177 and DUT-6 (MOF-205), respectively (see Fig. 5).

**Table 4** Selected bond lengths [Å] and bond angles [°] of MIL-53 (152 atoms in unit cell) optimized at DFTB level.

| Bond Type | DFTB | Exp. |
|---|---|---|
| Al-Al | 3.41 | 3.31 |
| Al-O | 1.89 | 1.83-1.91 |
| O-C | 1.33 | 1.29, 1.30 |
| O-Al-O | 88.4, 91.2 | 88.6, 89.8 |
| Al-O-Al | 128.9 | 124.9 |
| Cell paramet. | a=12.46 | a=12.18 |
| | b=17.32 | b=17.13 |
| | c=13.65 | c=13.26 |
| | $\alpha=\beta=\gamma=90$ | $\alpha=\beta=\gamma=90$ |

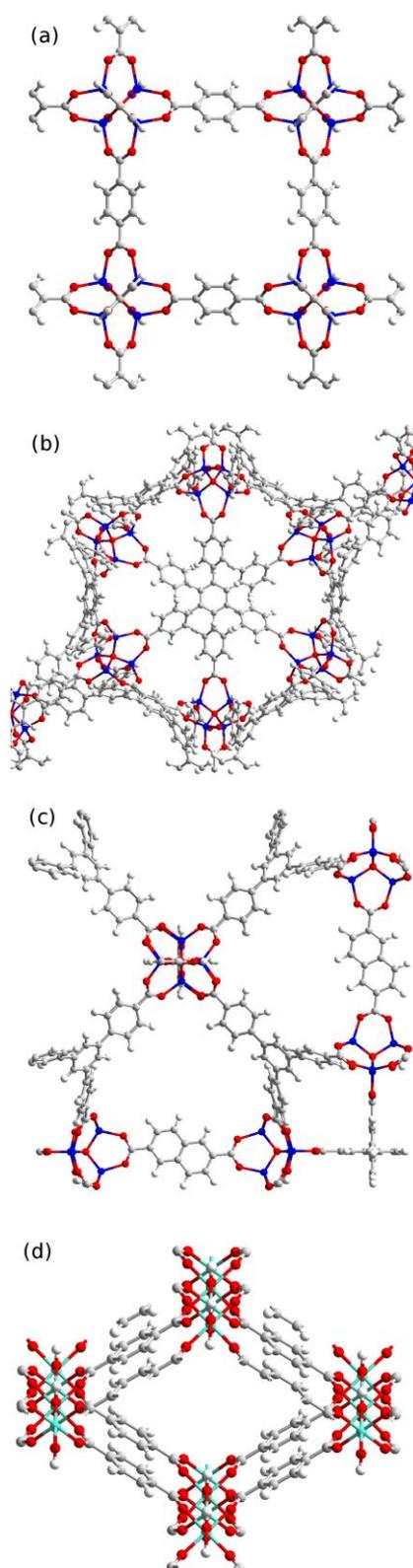

**Figure 5** The crystal structures in the unit cell representations of (a) MOF-5, (b) MOF-177, (c) DUT-6 (MOF-205), and (d) MIL-53. red-oxygen, grey-carbon, white-hydrogen, blue-metal atom (Zn, Al).

All three MOFs have different topologies due to the organic linkers, where the number of connections is varied or where two different linker types are present. MOF-5 is the most simple and it is the prototype in the isoreticular series (IRMOF-1) [70], it is a simple cubic topology with three-dimensional pores of the same size, and the linkers have only two connection points. In the case of MOF-177, the linker is represented by a triangular SBU, that means three connection points are present. This results in the topology with mixed (3,6)-connectivity [71]. DUT-6 (MOF-205) has a much more complicated topology due to two types of linkers. The first one (NDC) has just two connection points, while the second is the same as in MOF-177 with three connection points. One Zn-O-based connector is connected to two NDC and four BTB linkers. The topology is very interesting; all rings of the underlying nets are 5-fold, and it forms a face-transitive tiling of dodectahedra and tetrahedra with a ratio of 1:3 [75].

The bond lengths in all three MOFs are nearly the same: Zn-Zn is around 3.3 Å, Zn-O around 2.0 Å, and C-O around 1.3 Å. These are generally overestimated comparing with the experimental data by less than 5%. The Zn-Zn distance is much longer than that of Cu-Cu in CuBTC, as there is no direct bond distance between metal atoms. On the other hand, the C-O bond is slightly shorter comparing with CuBTC, indicating stronger binding.

MIL-53 is a MOF structure with one-dimensional pores. The framework is built up of corner-sharing Al-O-based octahedral clusters interconnected with BDC linkers. The linkers have again two connection points. MIL-53 shows reversible structural changes dependent on the guest molecules [74]. It undergoes the so-called breathing mode depending on the temperature and the amount of adsorbed molecules.

In this case also the bond lengths and bond angles are slightly overestimated comparing with the experimental structures, but the error does not exceed 3%.

### 4 Mechanical Properties

Due to the low mass density the elastic constants of porous materials are a very sensitive indicator of their mechanical stability. For crystal structures like MOFs, we have studied these by means of bulk modulus ($B$). $B$ can be calculated as a second derivative of energy with respect to the volume of the crystal (here unit cell).

The result shows that CuBTC has bulk modulus of 34.66 GPa, what is in close agreement with $B$=35.17 GPa obtained using force-field calculations [75]. Bulk moduli for the series of MOFs resulted in 15.34 GPa, 10.10 GPa and 10.73 GPa for MOF-5, -177 and DUT-6 (MOF-205), respectively. For MOF-5, $B$=15.37 GPa was reported from DFT/GGA calculations using plane-waves [76]. The results show that larger linkers give mechanically less stable structures, what might be an issue for porous structures with larger voids. For MIL-53 we have obtained the largest bulk modulus of 53.69 GPa, keeping the angles of the pore fixed.

### 5 Conclusions

We have validated the DFTB method with self-consistent charge and London dispersion corrections for various types of metal-organic frameworks. The method gives excellent geometrical parameters compared to experiment and, for small model systems, also in comparison with density-functional theory calculations. Importantly, this statement holds not only for catalytically inactive MOFs based on the $Zn_4O(CO_2)_6$ octahedron and organic linkers, which are important for gas adsorption and separation applications, but also for catalytically active HKUST-1 (CuBTC). This is remarkable as the Cu atoms are in the $Cu^{2+}$ state in this framework. DFTB parameters have been generated and validated for Cu, and the electronic structure contains one unpaired electron per Cu atom in the unit cell, which makes the electronic description technically difficult, but manageable within the SCC-DFTB method. SCC-DFT performs well for the frameworks themselves as well as for adsorbed CO and water molecules.

We finally conclude that we have now a high-performing quantum method available to study various classes of MOFs of unit cells up to 10000 atoms, including structural characteristics, the formation and breaking of chemical and coordination bonds, for the simulation of diffusion of adsorbate molecules or lattice defects as well as electronic properties. The parameters can be downloaded from the DFTB.org website.

**Acknowledgements**

Financial support by Bulgarian National Science Fund (Contracts DSVP 02/1 and DCVP 02/2), Deutsche Forschungsgemeinschaft (DFG) within the Priority Program Metal-Organic Frameworks (SPP 1362), the European Research Council (ERC StG 56962) the European Commission (HYPOMAP NMP3-SL-2008-233482, QUASINANO MC-IAPP-2009-251149) and the DAAD (joint Bulgarian-German program PROBEMOF) is gratefully acknowledged.